\documentclass[aps,prl,reprint,showpacs,twocolumn,superscriptaddress]{revtex4}

\usepackage[dvips]{graphicx}
\usepackage{dcolumn}
\usepackage{bm}
\usepackage{xspace}
\usepackage{multirow}
\usepackage{natbib}
\usepackage{color}

\begin{document}

\preprint{}
\title{Absence of a Proximity Effect in a Topological Insulator on a Cuprate Superconductor: Bi$_2$Se$_3$/Bi$_2$Sr$_2$CaCu$_2$O$_{8+\delta}$}
\author{T. Yilmaz}
\affiliation{Department of Physics, University of Connecticut, Storrs, Connecticut 06269, USA\\}
\author{I. Pletikosi\'{c}}
\affiliation{Condensed Matter Physics and Materials Science Department, Brookhaven National Lab, Upton, New York 11973, USA}
\affiliation{Department of Physics, Princeton University, Princeton, NJ 08544, USA}
\author{A. P. Weber}
\affiliation{National Synchrotron Light Source, Brookhaven National Lab, Upton, New York 11973, USA}
\author{J. T. Sadowski}
\affiliation{Center for Functional Nanomaterials, Brookhaven National Lab, Upton, NY 11973, USA}
\author{G. D. Gu}
\affiliation{Condensed Matter Physics and Materials Science Department, Brookhaven National Lab, Upton, New York 11973, USA}
\author{A. N. Caruso}
\affiliation{Department of Physics, University of Missouri-Kansas City, Kansas City, Missouri 64110, USA}
\author{B. Sinkovic}
\affiliation{Department of Physics, University of Connecticut, Storrs, Connecticut 06269, USA\\}
\author{T. Valla}
\email{valla@bnl.gov}
\affiliation{Condensed Matter Physics and Materials Science Department, Brookhaven National Lab, Upton, New York 11973, USA}
\date{\today}

\begin{abstract}
Proximity-induced superconductivity in a 3D topological insulator represents a new avenue for observing zero-energy Majorana fermions inside vortex cores. Relatively small gaps and low transition temperatures of conventional $s$-wave superconductors put the hard constraints on these experiments. Significantly larger gaps and higher transition temperatures in cuprate superconductors might be an attractive alternative to considerably relax these constraints, but it is not clear whether the proximity effect would be effective in heterostructures involving cuprates and topological insulators.
Here, we present angle-resolved photoemission studies of thin Bi$_2$Se$_3$ films grown \textit{in-situ} on optimally doped Bi$_2$Sr$_2$CaCu$_2$O$_{8+\delta}$ substrates that show the absence of proximity-induced gaps on the surfaces of Bi$_2$Se$_3$ films as thin as a 1.5 quintuple layer. These results suggest that the superconducting proximity effect between a cuprate superconductor and a topological insulator is strongly suppressed, likely due to a very short coherence length along the $c$-axis, incompatible crystal and pairing symmetries at the interface, small size of the topological surface state\rq{}s Fermi surface and adverse effects of a strong spin-orbit coupling in the topological material.

\end{abstract}
\vspace{1.0cm}


\maketitle
Superconductivity in 3D topological insulators (TI) represents a promising avenue for observing zero-energy Majorana fermions \cite{Fu2008,Hasan2010,Qi2011}. The characteristic energy window, open in the vortex cores for such experiments is $\Delta^2/E_F$, where $\Delta$ represents the superconducting gap at the Fermi surface of topological surface state (TSS) and $E_F$ is the Fermi energy relative to the Dirac point. Due to the extremely weak electron-phonon coupling on the TSS and its inherent 2D character, the intrinsic superconductivity in the TSS would be highly unlikely and it would have to be induced by proximity to a superconducting bulk \cite{Pan2012,Hochst2005,Pan2013}. So far, bulk superconductivity has been observed only in Cu intercalated Bi$_2$Se$_3$ and in Bi$_2$Te$_3$ under pressure, with very low ($T_c<5$ K) transition temperatures in both systems \cite{Hor2010,Kriener2011,Zhang2011}. Such a low $T_c$ implies a sub-meV gap and the level spacing of $\sim10^{-8}$eV inside vortex cores, making the detection of Majorana modes virtually impossible in these materials. An alternative approach is to use artificial heterostructures, either lateral, or layered, involving various superconductors and topological insulators in which the superconductivity in topological states would be induced by a superconducting proximity effect near their interfaces \cite{Zhang2011a,Zareapour2012,Cho2013,Qu2012,Veldhorst2012,Williams2012,Yang2012}. 
Cuprate superconductors have the highest transition temperatures and largest gaps and might look as ideal candidates, but it is not clear if the proximity effect would be efficient at interfaces involving materials with incompatible crystal and pairing symmetries \cite{Qi2011}. In addition, it is not clear if a superconductivity induced by proximity to a $d$-wave superconductor, with its intrinsic zero-energy modes in vortex cores, would allow detection of Majorana fermions. 
The recent study of a thin Bi$_2$Se$_3$ film on a Bi$_2$Sr$_2$CaCu$_2$O$_{8+\delta}$ (BSCCO) substrate has indicated that superconductivity could be induced in the TSS, even for relatively thick films of 6-7 quintuple layers (QL), in which the TSS is fully formed and does not overlap with the interface state \cite{Wang2013}. Surprisingly, the gap was found to be fairly large ($\sim$ 15 meV) and isotropic. If true, these findings would be very promising for prospects of detecting the Majorana modes as this would increase the level spacings in the vortex cores by 3-4 orders of magnitude, with a potential for even more if the $E_F$ could be lowered by controlled filling of the TSS. However, there are some very inconsistent observations in that study that call for caution: (1) the ARPES study was done on \textit{ex-situ} prepared films, requiring a delicate recovery procedure, while the gap was not observed in scanning tunneling microscopy (STM) on \textit{in-situ} grown films; (2) the gap was detected only in the TSS and not in the quantum well states (QWS) (precursors of the bulk conduction band), even though the spatial extension of the QWS throughout the film thickness would favor the larger proximity effect in them. Actually, the proximity effect could only be propagated to the top surface by these QWS as the TSS is otherwise completely decoupled from the interface; (3) the absence of the coherent quasi-particle peak in the TSS with such a large gap would suggest that the shift might be caused by some artifacts and not by a superconducting gap \cite{In}.

Here, we study the \textit{in-situ} grown Bi$_2$Se$_3$ films on optimally doped BSCCO substrates, in the thickness range from 0.5 to 12 QL. We find no evidence of proximity-induced gaps at the Fermi surface of the top TSS within our experimental uncertainty of $\sim$ 1 meV. This suggests that the proximity effect in heterostructures involving the interfaces between a $d$-wave superconductor and TI layers is strongly suppressed, likely because of a short coherence length along the $c$-axis in a cuprate superconductor \cite{Palstra1988}, a small size of the Fermi surface of the TSS and possibly due to adverse effects of a strong spin-orbit coupling (SOC) in the TI layer on the $d$-wave pairing near the interface \cite{Grimaldi1999}.

\begin{figure}[htbp]
\begin{center}
\includegraphics[width=8.75cm]{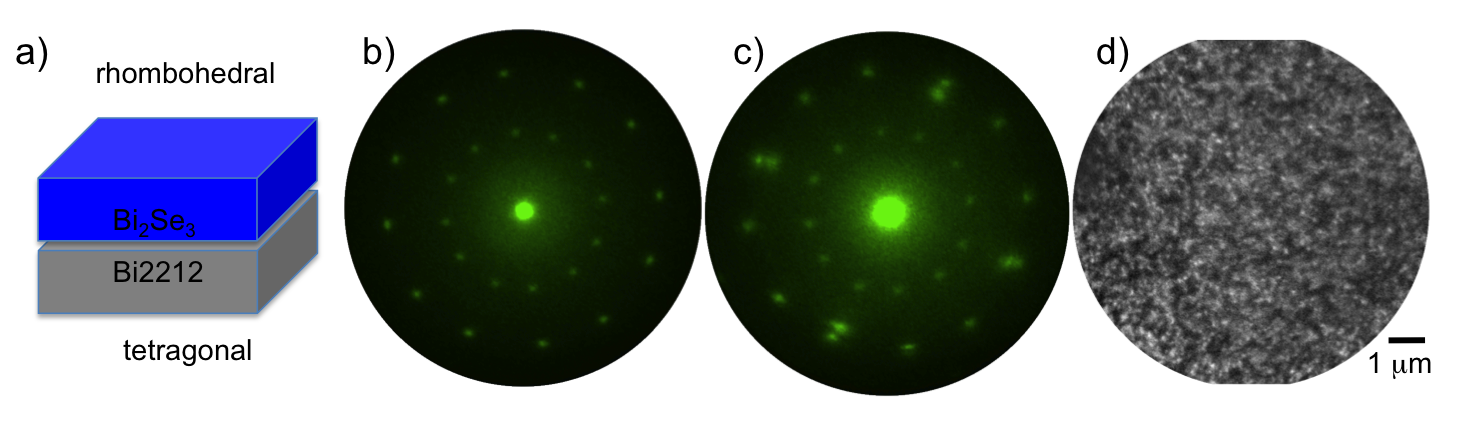}
\caption{(Color on line) Structure of a Bi$_2$Se$_3$ film on a BSCCO substrate. (a) Schematic view of the studied structures (b) $\mu$-LEED image of a 10 QL Bi$_2$Se$_3$ film at 350 K. Two equally populated domains, rotated by 30$^{\circ}$ relative to each other, contribute to diffraction intensity (c) $\mu$-LEED image at T=750K. Film has mostly desorbed and tetragonal pattern from the substrate becomes visible. (d) A real-space, dark-field LEEM image obtained from a diffraction spot corresponding to one of the domains. Bright features, representing the domains that contribute to the chosen diffraction spot, indicate the domain size in the sub-$\mu$m range. Images were taken at 20 and 18 eV in diffraction and real-space imaging mode, respectively.
}
\label{Fig1}
\end{center}
\end{figure}

Optimally doped ($T_c=91$ K) BSCCO single-crystals were grown by the traveling floating zone method. They were cleaved \textit{in-situ} and annealed to 420 K prior to the film growth in an MBE chamber with the base pressure of 10$^{-7}$ Pa, attached to the ARPES chamber with the base pressure 3$\times$10$^{-9}$ Pa. Bi$_2$Se$_3$ films were grown under Se-rich conditions using high purity elements (5N) in the resistively heated evaporators and a \lq\lq{}two step growth method\rq\rq{}: a single QL layer was grown by keeping the substrate at 420 K, while the additional material was deposited at 470 K. After the growth, films were kept at 470 K for additional 10 minutes, cooled to room temperature and transfered to the ARPES chamber, all in UHV.
The ARPES experiments were carried out on a Scienta SES-R4000 electron spectrometer at beamline U5UA at the National Synchrotron Light Source, in the photon energy range from 17 to 52 eV. The analyzer is equipped with two Mott detectors and can work in spin-resolved (SR) mode. The total instrumental energy resolution in the ARPES mode was $\sim$ 6 meV at 17 eV and $\sim$ 10 meV at 52 eV photon energy.  Angular resolution was better than $\sim 0.15^{\circ}$ and $0.4^{\circ}$ along and perpendicular to the slit of the analyzer, respectively. SR-ARPES data were recorded at room temperature, at 52 eV photon energy and with $\sim$ 40 meV energy and $\sim0.4^{\circ}$ angular resolution.
Low-energy electron microscope (LEEM) and micro-spot low-energy electron diffraction ($\mu$-LEED) measurements were performed in the ELMITEC LEEM III endstation at the NSLS beamline U5UA. The $\mu$-LEED patterns were obtained from the area of 2 $µ\mu$m in diameter.

The rhombohedral crystal structure of Bi$_2$Se$_3$ is composed of hexagonal (111) atomic planes stacked in  QL units with the in-plane lattice parameter $a=4.14$ \AA. The tetragonal crystal structure of BSCCO is made up of square (100) planes having in-plane lattice parameter $a=3.82$ \AA. Due to the mismatch of the in-plane lattice symmetry, there are two possible orientations of (111) planes on top of (100) by matching of atomic rows along [100] or [010] direction of BSCCO with the [221] direction of the Bi$_2$Se$_3$ (111) layers. This would result in two Bi$_2$Se$_3$ domains with crystalline orientation effectively rotated by 30$^{\circ}$. We have performed $\mu$-LEED measurements on 10 QL Bi$_2$Se$_3$ film grown on BSCCO. As seen in the Fig.1(b) we indeed observe two hexagonal LEED patterns rotated by 30$^{\circ}$ with respect to each other, indicating presence of two distinct crystallographic domains. Equal intensity of the two patterns indicates their equal population under the e-beam spot.
Figure \ref{Fig1}(c) shows $\mu$-LEED pattern after annealing at T=750 K, which causes significant removal of the Bi$_2$Se$_3$ film and partial exposure of BSCCO surface. Thus, we note the appearance of additional, square diffraction pattern attributed to BSCCO (100) surface. The orientation of two types of Bi$_2$Se$_3$ domains on top of the BSCCO is evident from alignment of the (10) LEED spots of hexagonal and square patterns, consistent with idea of row matching of the two lattices described above.
In Figure \ref{Fig1}(d) we show the real space dark-field LEEM image using the diffraction beam of one of the Bi$_2$Se$_3$ domains. The white-gray contrast suggest that the domains are smaller than 1 $\mu$m. This also indicates that the ARPES data presented below will average over these two types of domains.

\begin{figure*}[htb]
\begin{center}
\includegraphics[width=14cm]{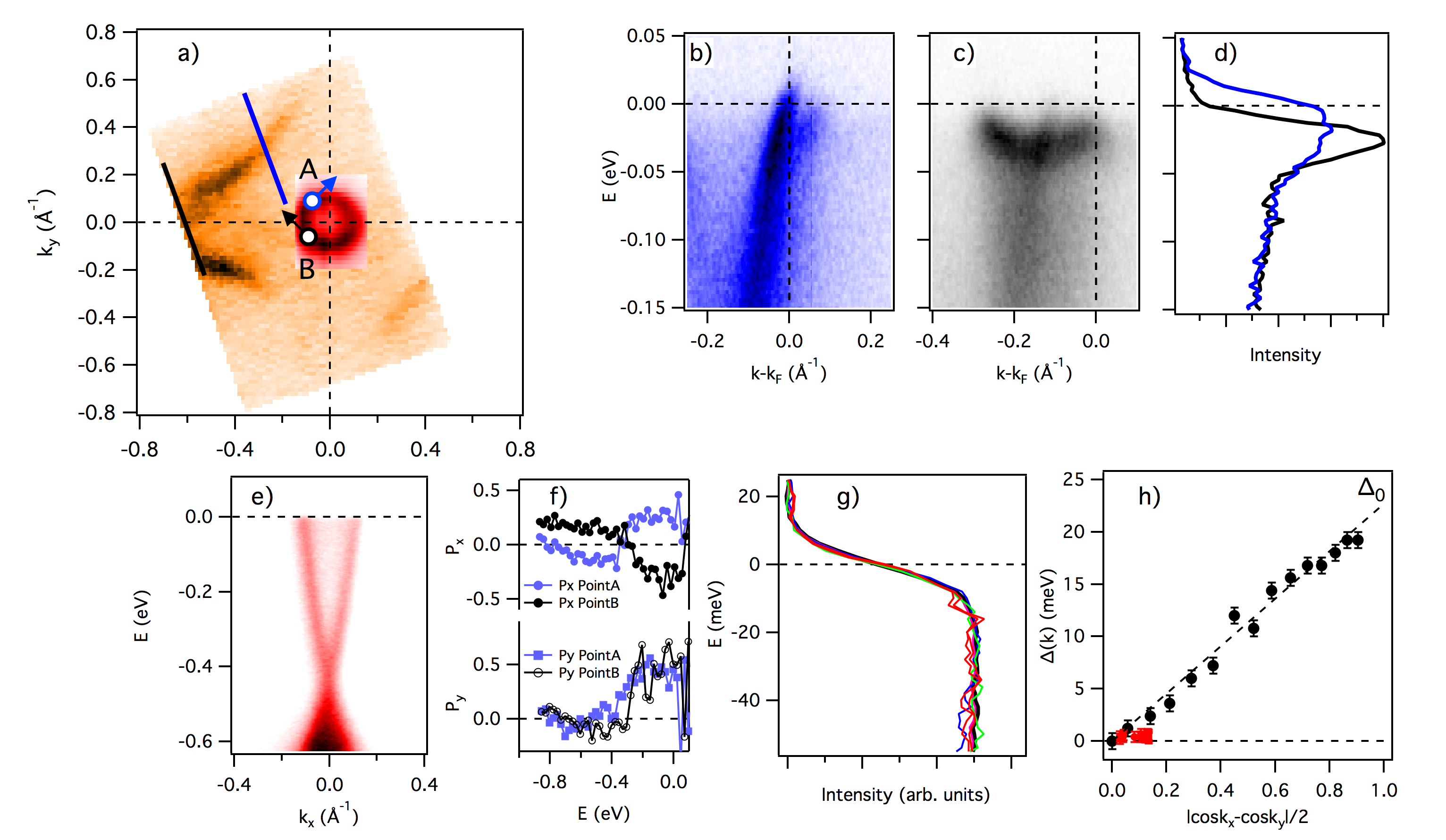}
\caption{ (Color on line) Electronic structure of the BSCCO substrate and the 10 QL Bi$_2$Se$_3$ film grown on it. (a) ARPES intensity from a BSCCO sample that served as a substrate for a 10 QL thick Bi$_2$Se$_3$ film at E=0 as a function of the in-plane momentum. Gold false-color scale corresponds to BSCCO (before the film growth), while red false-color scale corresponds to the film\rq{}s Fermi surface. Circles A and B represent points where SR-ARPES data were recorded, while the arrows represent the spin polarization. (b) and (c) ARPES spectra along the solid blue (black) line in (a), taken at T=30 K, representing the nodal (anti-nodal) region of BSCCO. (d) The EDC curves taken at the Fermi wave-vector from the nodal (anti-nodal) region of the BSCCO Fermi surface, shown as the blue (black) line. (e) ARPES intensity obtained from a 10 QL thick Bi$_2$Se$_3$ film at 30 K, showing a conical dispersion of TSS. (f) The in-plane components of spin polarization measured at points A and B on the Bi$_2$Se$_3$ Fermi surface. (g) EDC curves from several points on the circular Fermi surface of the TSS shown in (a), recorded at T=30 K. Thick black curve corresponds to the spectrum from the Au reference. (h) The momentum dependence of the BSCCO gap (black circles), showing a characteristic $d$-wave shape. Red points correspond to the Bi$_2$Se$_3$ film data. Note the very limited span on the $|$cos$k_x$-cos$k_y|/2$ scale, due to a small size of the Bi$_2$Se$_3$ Fermi surface. The spectra of pristine BSCCO substrate were recorded with 30 eV photons, while the Bi$_2$Se$_3$ film was measured at 50 eV photon energy. 
}
\label{Fig2}
\end{center}
\end{figure*}

Figure \ref{Fig2} shows ARPES measurements of the electronic structure of the BSCCO substrate prior and after depositing 10 QL thick Bi$_2$Se$_3$ film. The BSCCO crystal was cleaved \textit{in-situ} and annealed for $\sim$ 15 minutes to 420 K prior the measurements to check for the effects of elevated temperatures, required for the film growth, on the oxygen content in the near surface region. The size of the Fermi surface (panel (a)) and the superconducting gap measured at T=30 K indicate that the annealing process does not alter the electronic structure of the substrate. Panels (b-d) show the low temperature spectra from the nodal and anti-nodal regions of the Fermi surface and the gap values typical for an optimally doped BSCCO \cite{Damascelli2003,Valla2006a}. The complete $k$-dependence of the superconducting gap (panel (h)) shows a usual $d$-wave shape, with the $\Delta_0\approx 25$ meV. The gap magnitude is extracted from the leading edge of energy distribution curves (EDC) obtained through the integration over the narrow momentum region around the corresponding $k_F$. This method gives slightly smaller amplitudes than those obtained from positions of the quasi-particle peaks, but is shown to be less sensitive to the experimental resolution \cite{Valla2006a}. 

After depositing 10 QL thick Bi$_2$Se$_3$ film on a well-characterized substrate, we see a well developed cone (Fig. \ref{Fig2}(e)) of the TSS with no gap at the Dirac point and a circular Fermi surface (panel (a)), slightly larger than in a typical single crystal \cite{Pan2012}. The state also shows a fully developed helical spin texture, as shown in Fig. \ref{Fig2}(f). We note that features from the substrate are not visible any more due to the very short probing depth in the ARPES experiments. 
Figure \ref{Fig2}(g) shows the low energy region of EDCs from several different points on the TSS Fermi surface. None of the EDCs show a gap or a coherence peak that would indicate that the system is in a superconducting state. We therefore conclude that the proximity-induced gap on the surface of a 10 QL Bi$_2$Se$_3$ film is either non-existing, or it is beyond our detection limits, $\Delta(k)<0.5$ meV, in this particular case. This is not surprising as it is well known that the $c$-axis coherence length in BSCCO is extremely short, $\xi_c< 0.5$ nm, and the top surface in 10 QL film is already around 10 nm away from the nearest Cu-O plane \cite{Palstra1988}. Also, the TSS is relatively well localized to the surface, with the zero overlap with the interface region and a penetration of surface quasi-particles into the superconducting region of a junction essentially does not exist. 
Another reason for the lack of proximity effect might be a very small size of the TSS Fermi surface. From Fig.\ref{Fig2}(a) and (h) it is clear that the Fermi surface of the TSS is too small, $k_F=0.108$ \AA$^{-1}$, to acquire a significant $|$cos$k_x$-cos$k_y|/2$ magnitude. This might be a crucial factor if the proximity effect has to be propagated by the $d$-wave pairing. Although the Josephson tunneling is a $q=0$ process in a $d$-wave superconductor, and therefore allowed between the Fermi surfaces of different sizes, the induced gap will be smaller on a smaller Fermi surface according to the $|$cos$k_x$-cos$k_y|/2$ factor.

\begin{figure}[htbp]
\begin{center}
\includegraphics[width=8.5cm]{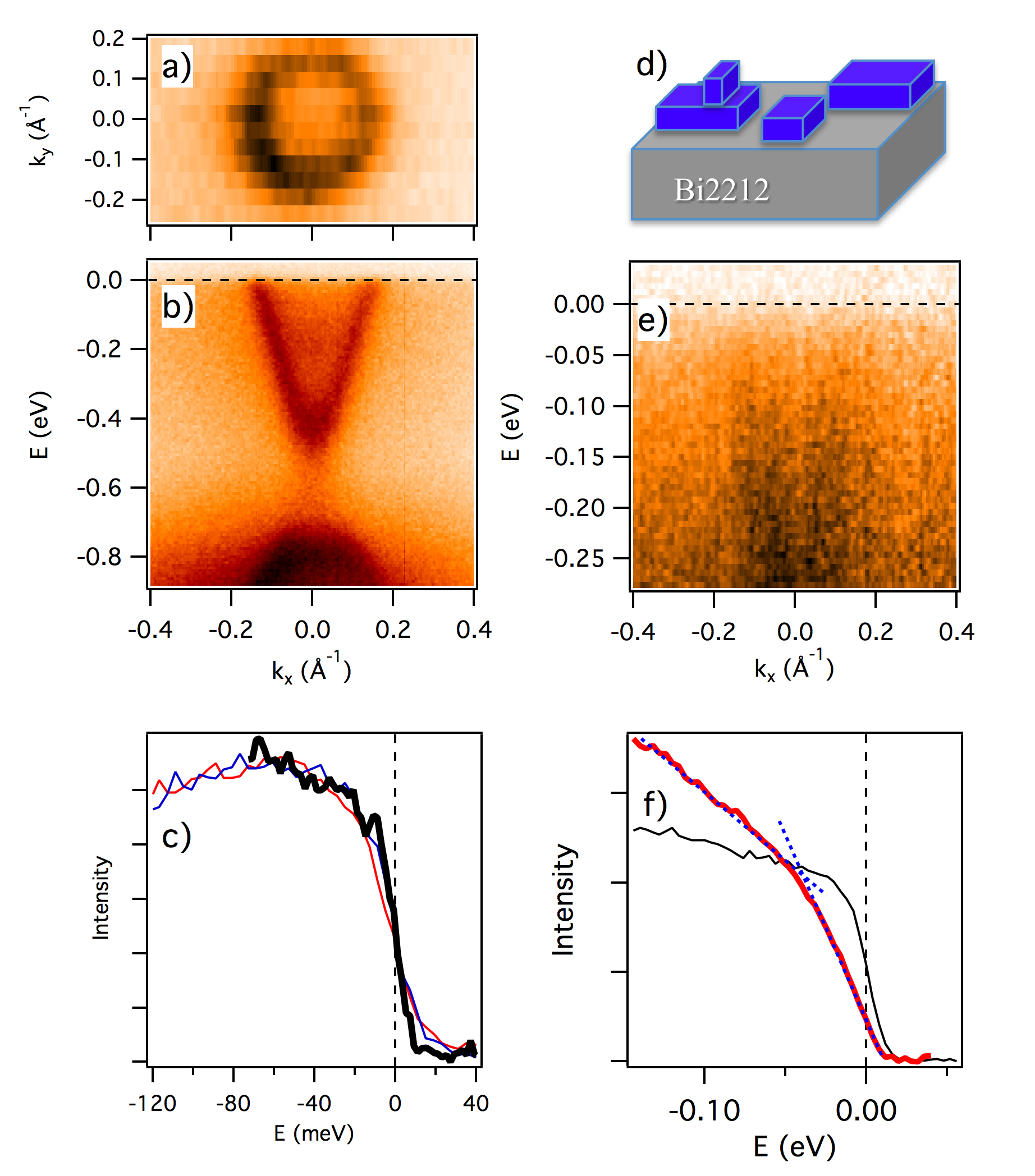}
\caption{(Color on line) Electronic structure of very thin Bi$_2$Se$_3$ films on a BSCCO substrate. (a) Femi Surface and (b) dispersion spectrum of a $\sim1.5$ QL thick Bi$_2$Se$_3$ film recorded at 30 K. (c) Several EDCs, corresponding to the spectra from $\sim1.5$ QL thick film shown in a) and b) and the EDC taken from an Au reference (thick black line). (d) A schematic representation of a sub-QL Bi$_2$Se$_3$ film. (e) ARPES spectrum from a $\sim0.5$ QL  Bi$_2$Se$_3$ film on a BSCCO substrate at 30 K. The precursor of TSS is visible near the $\Gamma$ point. (f) A characteristic EDC corresponding to the spectrum shown in e) (red line) and a spectrum from an Au reference (black line). The EDCs from the film do not show any $k$ dependence. The thin dashed lines indicate a change of slope at $\approx 40$ meV in the otherwise featureless lineshape.  
}
\label{Fig3}
\end{center}
\end{figure}

After negative results in a relatively thick film, we then turned to thiner films. We studied nominally 6, 4, 3, 2 and 1.5 QL thick films and did not find any evidence of proximity-induced gaps in any of them. Fig.\ref{Fig3} shows the results from two thinnest films studied here. Panels (a) and (b) show the Fermi surface and the dispersion of the surface state, while the panel (c) shows several EDCs from different points on the Fermi surface of a 1.5 QL thick Bi$_2$Se$_3$. The dispersion image, Fig.\ref{Fig3}(b), clearly shows a gap at the Dirac point, indicating the overlap of the surface and interface states. The Fermi surface, Fig.\ref{Fig3}(a),  is still reasonably sharp and slightly larger at this thickness ($k_F=0.145$ \AA$^{-1}$). Both the distance to the Cu-O planes in BSCOO ($\sim$1.5-2 nm) and the size of the Fermi surface are now more favorable to the proximity-induced superconductivity in the probed surface state. However, no gaps and no coherence quasi-particle peaks were detected in EDCs (panel (c)) within our experimental uncertainty, $\sim$1 meV in this particular case. We estimate the $d$-wave factor, $|$cos$k_x$-cos$k_y|/2$, for this Fermi surface to be around 0.18, which would result in $\sim$4.5 meV gap if no real-space tunneling was required (a direct contact of the top surface with the Cu-O plane). The finite distance and a short coherence length could further reduce that amplitude beyond our detection limits.

Finally, the only case where we do see a finite gap is in sub-QL Bi$_2$Se$_3$ films. Figures \ref{Fig3}(e) and (f) show the dispersion and an EDC obtained from a $\sim0.5$ QL thick film. As indicated in the schematic view, this film necessarily forms an inhomogeneous structure, made of 1-2 QL thick Bi$_2$Se$_3$ segments and exposed BSCCO areas. The substantial background and broad features in the spectrum shown in panel (e) indicate a high degree of disorder and scattering. The undeveloped Bi$_2$Se$_3$ state suggests that the Bi$_2$Se$_3$ segments might be too small for the coherent $k$-space picture. At this thickness, the BSCCO substrate is still probed, but its states have also become very broad. The EDCs now show a clear shift of the leading edge by $\sim$16 meV from the Fermi level. However, instead of a clear gap and a coherence peak, as seen in Fig.\ref{Fig2}(d), the spectrum shows a $k$-independent linear reduction in density of states at energies $|E|<40$ meV,  reflecting the linear density of states inside the $d$-wave gap in BSCCO \cite{Valla2006}. The lack of coherence peaks indicates that the state is phase-incoherent, probably due to the strong disorder and the negative influence of the film on the $d$-wave phase \cite{Emery1995}.  

In summary, the results presented here show no evidence of a proximity-induced superconducting gap on the surfaces of Bi$_2$Se$_3$ films as thin as a $\sim$1.5 QL, in stark contrast to the recent study by Wang \textit{et al} \cite{Wang2013}. 
We identify possible origins of a vanishing proximity effect in a very short $\xi_c$ in BSCCO, small Fermi surface of the TSS, incompatible crystal and pairing symmetries in the two materials and unfavorable effects of the strong SOC in Bi$_2$Se$_3$ on the $d$-wave pairing near the interface.  
Despite the negative results, this study gives important insights not only into the nature of low-energy electronic excitations in topological states, but also offers a better understanding of cuprate superconductors and processes involving the interfaces of these materials. It also offers suggestions for new experiments that could further test some of the proposed origins of a vanishing proximity effect in similar heterostructures. For example, it might be possible to increase the filling of the topological state by \textit{in-situ} doping to the point where the $d$-wave factor on its Fermi surface acquires a significant value \cite{Valla2012a}. Also, different materials, not necessarily TI, should be studied to test the role of the crystal structure of the interface and the strength of SOC in deposited material on the proximity effects near the BSCCO interface.

\begin{acknowledgments}
This work was supported by the US Department of Energy, Office of Basic Energy Sciences, contract
no. DE-AC02-98CH10886, ARO MURI program, grant W911NF-12-1-0461. 

\bibliographystyle{apsrev}

\begin{thebibliography}{25}
\bibitem{In} A Proper grounding of samples is crucial for the stability of the measured energy of photoelectrons on the scale $< 1$ meV. Typical clamping of BSCCO samples with Ta was proven to be insufficient as it often caused a poor ground connection and a finite, up to 10 meV energy shift, manifested as a gap in the ARPES spectra. To avoid this, we placed a thin In foil under each BSCCO sample that melted during the film growth and formed a perfect ground connection during the ARPES experiments. In was also found to greatly improve the thermal contact.
\end{thebibliography}

\end{acknowledgments}

\end{document}